\documentclass[aps,prb,twocolumn,superscriptaddress]{revtex4}
\usepackage{graphicx}
\usepackage{amsmath,epsfig}
\newcommand{\SFMO}{Sr$_2$FeMoO$_6$ }
\newcommand{\Tc}{T$_{\mathrm{c}}$}
\newcommand{\Sd}{\mathrm{f}}
\newcommand{\dn}{\downarrow}
\newcommand{\up}{\uparrow}
\newcommand{\tp}{t^{\prime}}

\begin{document}

\title{Theory of Half-Metallic Double Perovskites I: Double Exchange Mechanism}
\author{O. Nganba Meetei}
\affiliation{Department of Physics, The Ohio State University, Columbus, Ohio 43210, USA}

\author{Onur Erten}
\affiliation{Department of Physics, The Ohio State University, Columbus, Ohio 43210, USA}

\author{Anamitra Mukherjee}
\affiliation{Department of Physics, The Ohio State University, Columbus, Ohio 43210, USA}
\affiliation{Department of Physics and Astronomy, University of British Columbia, Vancouver, BC V6T 1Z1, Canada}

\author{Mohit Randeria}
\affiliation{Department of Physics, The Ohio State University, Columbus, Ohio 43210, USA}

\author{Nandini Trivedi}  
\affiliation{Department of Physics, The Ohio State University, Columbus, Ohio 43210, USA}

\author{Patrick Woodward}
\affiliation{Department of Chemistry, The Ohio State University, Columbus, Ohio 43210, USA}

\begin{abstract}
The double perovskite material \SFMO has the rare and desirable combination of a half-metallic ground state with 100\% spin polarization and ferrimagnetic \Tc$\simeq 420$K, well above room temperature. In this two-part paper, we present a comprehensive 
theoretical study of the magnetic and electronic properties of half metallic double perovskites. In this paper we present exact diagonalization calculations of the ``fast" Mo electronic degrees coupled to ``slow" Fe core spin fluctuations treated by classical Monte Carlo techniques.
From the temperature dependence of the spin-resolved density of states, we show that the electronic polarization at the chemical potential is proportional to magnetization as a function of temperature. 
We also consider the effects of disorder and show that excess Fe leaves the ground state half-metallic while anti-site disorder greatly reduces the polarization. 
In a companion paper titled ``Theory of Half-Metallic Double Perovskites II: Effective Spin Hamiltonian and Disorder Effects", we derive an effective classical spin Hamiltonian that provides a new framework for understanding the magnetic properties of half-metallic double perovskites including the effects of disorder.  
Our results on the dependence of the spin polarization on temperature and disorder has important implications for spintronics.
\end{abstract}

\maketitle

\section{Introduction}

Materials with half-metallic ground states in which conduction electrons are fully spin polarized and have ferromagnetic transition temperatures \Tc \, well above room temperature are very rare in nature. Only two families of materials, the double perovskites and the Heussler alloys, have exhibited this special combination of properties. Consequently, they hold the potential for tremendous advancements in the field of spintronics as spin injectors and tunneling magnetoresistance devices. 

Here, we focus on double perovskites which have generated considerable interest due to their close connections to ternary perovskites. The perovskite family is known to exhibit a wide variety of exotic properties including high-\Tc \, superconductivity, colossal magnetoresistance and ferroelectricity. Moreover, since these materials are derived from the same general family there is the potential to grow lattice-matched layered materials with different functional properties in each layer. Double perovskites with the general formula A$_2$BB$^{\prime}$O$_6$ is a composite of two different ternary perovskites ABO$_3$ and AB$^{\prime}$O$_3$ arranged in a 3D checker board pattern.  The additional flexibility of choosing two different transition metal ions in double perovskites opens up many new avenues of material exploration, like the juxtaposition of strong spin-orbit coupling and strong interaction by combining 5d and 3d transition metals. Already the range of properties span metals to band insulators, and multi-band Mott insulators, as well as ferromagnets, antiferromagnets, ferroelectrics, multiferroics, and spin liquids. \cite{ibarra,Kobayashi,Balents_DP1,Balents_DP2,nganba_2012}

The most widely studied double perovskite \SFMO (SFMO) has a half metallic ground state with a ferromagnetic transition temperature \Tc $\approx$ 420K (Ref.~[\!\!\citenum{Kobayashi,dd-sarma}]) which is well above room temperature. In spite of having a complex chemical structure, SFMO is a simple system to understand from a theoretical point of view. In contrast to ferromagnets like iron, there is a clear separation of the localized (Fe spins) and itinerant degrees of freedom (originating from Mo). Unlike the  manganites, SFMO has neither Jahn-Teller distortions nor competing magnetic ground states. Finally, in contrast to dilute magnetic semiconductors, disorder is not an essential aspect of the theoretical problem. Previous theoretical work on half-metallic double perovskites includes pioneering T=0 electronic structure calculations \cite{Kobayashi,dd-sarma}, model Hamiltonians analyzed using various mean-field theories \cite{chattopadhyay_2001, alonso_2003, brey_2006}, and two-dimensional (2D) simulations \cite{sanyal_2009}. 

In this paper, hereafter referred to as paper I, and its companion paper titled ``Theory of Half-Metallic Double Perovskites II: Effective Spin Hamiltonian and Disorder Effects", hereafter referred to as paper II [\!\!\citenum{paper2}], we expand on our recent work \cite{prl_2011} on the magnetic and electronic properties of SFMO. Several new results as well as important details which were omitted in our Letter\citep{prl_2011} are covered. Broadly, paper I focuses on the properties of the itinerant quantum electrons and their effect on magnetism, while paper II describes in detail the derivation of the effective spin Hamiltonian and the results obtained from it. 

The main results presented in paper I are: (1) Using a variational analysis, we obtain a phase diagram as a function of the parameters in the Hamiltonian and show that for the specific parameters of SFMO, it is firmly in the {\em ferrimagnetic} phase. 
(2) We present, for the first time, the temperature dependence of the spin resolved density of states. 
The electronic polarization decreases from 100\% at T=0 with increasing temperature and vanishes above \Tc. More importantly, it shows that the polarization at the chemical potential is proportional to the core spin magnetization as a function of temperature. This result is crucial because it allows us to infer electronic properties from magnetic properties obtained from the effective spin Hamiltonian. (3) Finally, we present the dependence of the polarization on disorder. In particular, we show that for Fe rich systems, the ground state remains half-metallic while anti-site disorder rapidly reduces polarization.  

Paper I is organized as follows. In Sec.~\ref{sec:hamiltonian}, we describe the generalized double exchange Hamiltonian used to study SFMO. In Sec.~\ref{sec:variational_analysis}, we present a variational analysis that describes the dependence of ground state magnetic properties on Hamiltonian parameters. For parameters relevant to SFMO, the ground state is deep in the ferrimagnetic phase. The results of a perturbative spin wave analysis are discussed in Sec.~\ref{sec:sw}. We show that the spin stiffness, which sets the scale for magnetic \Tc, is two orders of magnitude smaller than the electronic energy scale, allowing us to separate the itinerant degrees of freedom from the localized spins in the spirit of the Born-Oppenheimer approximation.  The temperature dependence of the spin resolved density of states (DOS) is presented in Sec.~\ref{sec:DOS}. Finally, in Sec.~\ref{sec:disorder}, we present the effect of disorder on electronic polarization and conclude with some remarks about future directions in Sec.~\ref{sec:conclusion}.

\section{Model Hamiltonian} \label{sec:hamiltonian}

\begin{figure}[t!]
\vspace{0.2cm}
\centerline{
\includegraphics[width=7cm]{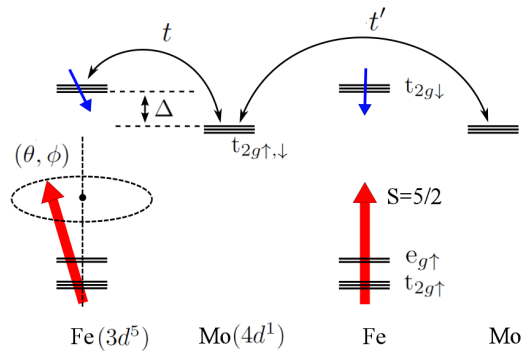}
}
\vspace{.2cm}
\caption{
Schematic showing energy levels at the transition metal sites in two unit cells (formula units) of SFMO. The Fe sites have localized $S=5/2$ core spins (red arrows), treated as classical vectors with orientation ($\theta,\phi$). The up and down sectors on the Fe site are split by a combination of the Hund's coupling $J_H$ and onsite Hubbard term $U$. The parameters $t$, $t^\prime$, and $\Delta$ of the Hamiltonian (\ref{eq:quantum_hamiltonian}), governing the dynamics of the itinerant electrons (blue arrows) in $t_{2g}$ orbitals, are also shown.}
\label{fig:levels}
\end{figure}

SFMO can be well understood in terms of a generalized double exchange model \cite{chattopadhyay_2001, alonso_2003, brey_2006,sanyal_2009,prl_2011, Aligia_2001, Petrone_2002}. The large Hund's coupling on Fe$^{3+}$ ($3d^5$) leads to a local $S=5/2$ \emph{core} spin.
Since locally all the spin up states on Fe are occupied, the only channel for the $4d^1$ electron on Mo$^{5+}$ to delocalize is by hopping from one Mo site to the next via the unoccupied Fe down states. This naturally leads to an antiferromagnetic coupling between the core Fe spins and the itinerant Mo electrons due to Pauli's exclusion principle. 
The conduction band is formed by hybridization of the Fe $t_{2g\dn}$ and Mo $t_{2g}$ orbitals via oxygen. Symmetry dictates that $d_{\alpha\beta}$ electrons delocalize only in the ($\alpha,\beta$) plane \cite{harris_2004} where $\alpha, \beta = x,y,z$. The model Hamiltonian describing SFMO is 
\begin{eqnarray}
H&=& -t\sum_{\langle i,j\rangle,\sigma}
(\epsilon_{i\sigma}d^{\dagger}_{i\downarrow}c_{j\sigma}+h.c.)
\cr
&&
-t^{\prime}\sum_{\langle j,j^\prime \rangle,\sigma}c^{\dagger}_{j\sigma}c_{j^\prime \sigma}
+ \Delta \sum_{i}d^{\dagger}_{i\downarrow}d_{i\downarrow}
\label{eq:quantum_hamiltonian}
\end{eqnarray}
where $d_i \, (c_i)$ denotes the fermion operator on the $i^{th}$ Fe (Mo) site. The electronic spin on Fe site is quantized along the direction of the local spin, whereas on Mo site the quantization is along a global z-axis. $t$ is the nearest neighbor Fe-Mo hopping amplitude, $\tp$ is the direct hopping amplitude between Mo sites and $\Delta$ is the charge transfer energy between Fe $t_{2g\downarrow}$ and Mo $t_{2g}$ states. In the global frame of the Mo spins, the orientation of the  $i^{th}$ Fe core spin is given by $(\theta_i,\phi_i)$ and it determines the effective hopping amplitude between Fe and Mo sites through $\epsilon_{i\up}$ and $\epsilon_{i\dn}$ which are defined as
\begin{eqnarray}
\epsilon_{i \uparrow} &=& - \sin(\theta_i/2)\exp(i\phi_i/2) \nonumber \\
\epsilon_{i \downarrow} &=& \cos(\theta_i/2)\exp(-i\phi_i/2)
\end{eqnarray}
A schematic of the level structure is shown in Fig.~\ref{fig:levels}. In pure SFMO, we ignore direct Fe-Fe hopping and Fe-Fe superexchange because the Fe sites are far apart \cite{scaling_of_Jex} and the spatial extent of the 3d orbitals is much smaller compared to 4d orbitals. However, these can be important in the presence of disorder since two Fe sites can be right next to each other. 

At first glance, the double exchange model for double perovskites looks like the antiferromagnetic Kondo lattice model \cite{fazekas}. However, there are significant differences: In the case of the antiferromagnetic Kondo lattice, the local moments are quantum degrees of freedom whereas in SFMO the local moments are large ($S=5/2$) and treated classically. Another consequence of the large local moments in SFMO is that they cannot be completely screened by the available conduction electrons. Finally, as mentioned above, Pauli's exclusion principle is responsible for the antiferromagnetic coupling between the local moments and the itinerant electrons in SFMO. In contrast, the coupling in  Kondo lattice arises from antiferromagnetic exchange interaction and is typically small compared to the band width.

\section{Variational Analysis}\label{sec:variational_analysis}

\begin{figure}[!t]
\vspace{.2cm}
\centerline{
\includegraphics[width=8.5cm]{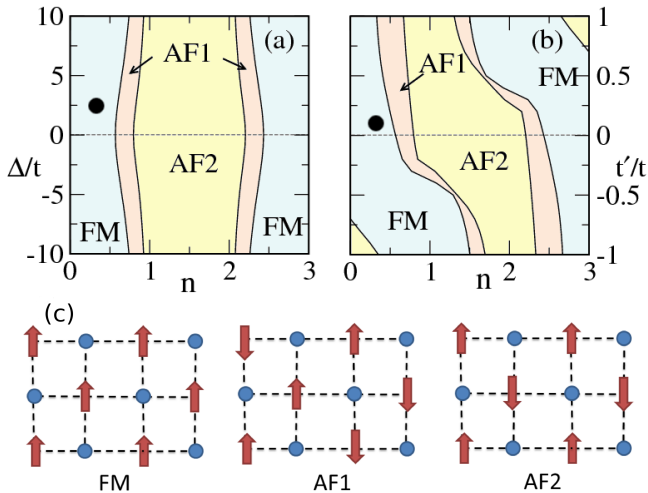}
}
\vspace{.2cm}
\caption{Variational calculation: (a) $\Delta$-n magnetic phase diagram for t$^\prime$=0; here $\Delta$ is the charge transfer offset between the Mo $t_{2g}$ orbitals and the Fe $t_{2g\dn}$ orbitals, and $n$ is the filling.
(b) t$^\prime$-n phase diagram for $\Delta$=0. Solid lines are first order phase boundaries and the black dot indicates SFMO parameters. Schematic of the variational magnetic phases,
ferromagnet (FM),  stripe antiferromagnet (AF1), and usual N{\'e}el antiferromagnet (AF2) are shown in (c). The red arrows indicate Fe spins and blue dots indicate Mo sites.}
\label{fig:phase_diagram}
\end{figure}

We begin by studying the T=0 properties of the quantum Hamiltonian in Eq.~\eqref{eq:quantum_hamiltonian}.  As a function of Hamiltonian parameters, we explore the relative stability of various magnetic phases shown in Fig.~\ref{fig:phase_diagram}(c): ferromagnetic (FM), stripe antiferromagnet (AF1) and N\'{e}el antiferrmomagnet (AF2). 
Note that the FM phase has induced moments on Mo sites which are aligned in the opposite direction. So, technically, it is a ferrimagnetic state, but we will focus on the core spins and refer to this state as FM. We use the same nomenclature in paper II also.
 In Fig.~\ref{fig:phase_diagram}(a), we present the $\Delta-n$ phase diagram for $\tp=0$, where $0\leq n \leq 3$ is the electron filling. SFMO corresponds to $n=1/3$. Part (b) shows the $\tp-n$ phase diagram for $\Delta$=0. The phase boundaries are determined by the relative energy of conduction electrons in the various variational spin backgrounds. For the FM state, the energy is calculated by integrating over the filled states in the band structure shown in Fig.~\ref{fig:DOS}(a). Details for the AF1 and AF2 phases are provided in Appendix \ref{appen:variational}. Our result in Fig.~\ref{fig:phase_diagram}(a) is consistent with previous calculations \cite{sanyal_2009}.

The scale for the magnetic \Tc \, is set by $t$, and we will show in paper II that choosing $t = 0.27$ eV, consistent with band structure calculations \cite{dd-sarma}, leads to the experimental \Tc$\approx 420$K of SFMO. For the other Hamiltonian parameters, we use  $\tp / t = 0.1$ and $\Delta /t = 2.5$ which are also in agreement with band structure calculations \cite{dd-sarma}. It is clearly seen from Fig.~\ref{fig:phase_diagram}(a) and (b) that, for the parameters relevant to SFMO (indicated by black dots), the ground state is deep inside the FM phase. This justifies our claim that SFMO does not have competing magnetic phases.

It is, however, interesting to note that t$^\prime$ has a major role in determining the phase boundaries. This is due to the fact that the effect of $t^\prime$ is very different in the different phases. For $t^\prime =0$, the AF1 phase has one dimensional bands and increasing $\vert t^\prime \vert$ introduces two dimensional hopping which changes the nature of the bands dramatically. However, in the case of FM and AF2 phases, the bands are two dimensional in nature even for $t^\prime =0$ and, therefore, the effect of $t^\prime$ is not as strong as in the AF1 phase.
The $\tp<0$ region of our phase diagram can be mapped on to the $\tp>0$ region by the symmetry of the Hamiltonian: $E(t^{\prime} , \Delta ,n)\equiv E(-t^{\prime}, -\Delta, 3-n)$.

On the other hand, $\Delta$ has very little effect on the T=0 phase diagram because its effect is very similar in all the phases.  We, however, expect increasing $\Delta$ to significantly reduce \Tc. These phase diagrams are useful in guiding materials search with optimized parameters.

\section{Spin Waves}\label{sec:sw}

\begin{figure}[t]
\centering
\includegraphics[width=5cm]{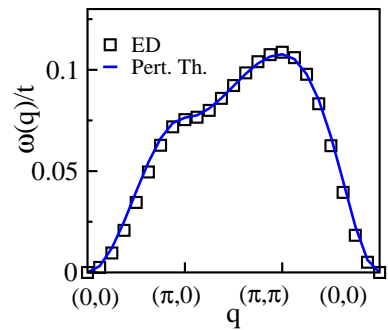}
\caption{Comparison of spin wave dispersion from perturbation theory [Eq.~\eqref{eq:sw_dispersion}] with that obtained from exact diagonalization of Eq.~\eqref{eq:quantum_hamiltonian} in a spin wave background. We find excellent agreement with no fitting parameters. Also note that the energy scale of magnetic interactions is two orders of magnitude smaller than the electronic band width ($\approx 8t$). }
\label{fig:sw}
\end{figure}

\begin{figure*}[t]
\centerline{
\includegraphics[width=17cm]{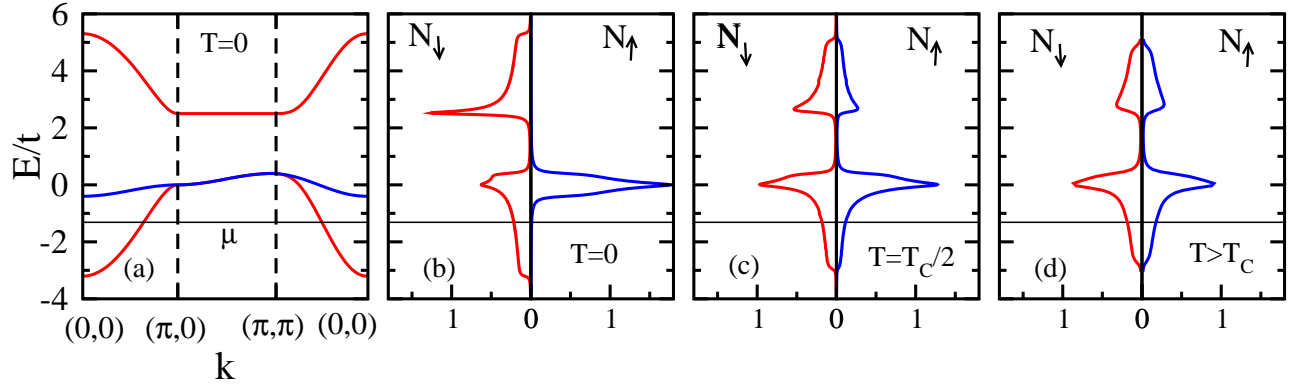}
}
\caption{(a) Electronic polarized band structure of FM ground state (blue indicates spin down and red, spin up). At SFMO filling, only the lowest spin-down band is occupied, thereby confirming the half-metallic ground state. (b), (c) and (d) shows the spin resolved density of states at T=0, T$\approx$\Tc/2 and T$>$\Tc respectively. The black horizontal line indicates chemical potential at SFMO filling.}
\label{fig:DOS}
\end{figure*} 

In this section, we calculate the spin wave dispersion and the spin stiffness of SFMO using a perturbative analysis. At low temperatures, the core Fe spins fluctuate about the fully magnetized ferromagnetic state and these fluctuations affect the mobile electrons. To lowest order, we can view the core spin fluctuations as generating spin wave configurations that are static on the time scale of the electronic degrees of freedom. This separation of time scales for the core spins and the itinerant electrons will be justified \textit{a posteriori}, and plays an important role for later results.  

The classical core spins in a frozen spin wave can be described as
\begin{align} \label{eq:spin_wave}
  S_i^z &= \cos \theta \nonumber \\
  S_i^x &= \sin \theta \cos( \mathbf{q \cdot r_i}) \nonumber \\
  S_i^y &= \sin \theta \sin( \mathbf{q \cdot r_i})
\end{align}
where {\bf q} is the wave vector of the spin wave. We assume that the angle $\theta$ with respect to the quantization axis of the FM ground state is small and we explicitly calculate the corrections to the energy up to $\mathcal{O}(\theta^2)$. 
In Eq.~\eqref{eq:quantum_hamiltonian} the fermion operators on the Fe sites are described with respect to the {\em local} quantization axis. In the analysis here, for small angular perturbations of the Fe spins, it is convenient 
to redefine the creation(annihilation) operators on Fe sites in the same global frame as the operators on Mo sites. We choose
\begin{align} \label{eq:global_fe_operator}
  d_{i\dn}^{\dagger} &= \cos\left(\theta/2\right)\Sd_{i\dn}^{\dagger} - \sin\left(\theta/2\right)e^{-i\mathbf{q \cdot 	r_i}}\Sd_{i\up}^{\dagger} 
\end{align}
where $\Sd_{i\sigma}^{\dagger}$($\Sd_{i\sigma}$) is the creation(annihilation) operator in the global frame for an electron with spin $\sigma$ on the $i^{th}$ Fe ion. The Hamiltonian in Eq.~\eqref{eq:quantum_hamiltonian} can now be rewritten in terms of these new operators. Keeping only terms up to $\mathcal{O}(\theta^2)$, we get 
\begin{equation} \label{eq:pert_expansion}
  H = H_0 + \theta H_1 + \theta^2 H_2
\end{equation}
where 
\begin{align} 
  H_0 &= \Delta\sum_{i\sigma} \Sd_{i\sigma}^{\dagger}\Sd_{i\sigma} -t\sum_{\langle i,j \rangle} \left( \Sd_{i\dn}^{\dagger}c_{j\dn} + h.c. \right) \nonumber \\
      &-  t^{\prime} \sum_{\langle\langle i,j \rangle\rangle,\sigma} \left( c_{i\sigma}^{\dagger}c_{j\sigma} + h.c. \right)  \label{eq:H_0} \\       
  H_1 &= \frac{t}{2} \sum_{\langle i,j \rangle}\left( e^{-i\mathbf{q \cdot r_i}}\Sd_{i\up}^{\dagger}c_{j\dn} + e^{i\mathbf{q \cdot r_i}}\Sd_{i\dn}^{\dagger}c_{j\up} + h.c.\right) \label{eq:H_1} \\
  H_2 &= \frac{t}{4}\sum_{\langle i,j \rangle} \left( \Sd_{i\dn}^{\dagger}c_{j\dn} - \Sd_{i\up}^{\dagger}c_{j\up} + h.c. \right) \label{eq:H_2} 
\end{align}

Here $H_0$ is the unperturbed Hamiltonian, while $H_1$ describes the hybridization of spin up and spin down orbitals which is unique to the double perovskites. Finally, $H_2$ contains terms responsible for narrowing the spin down conduction band while allowing spin up electrons to delocalize. The details of the calculation are given in Appendix \ref{appen:perturbation}.
  
One subtlety that is worth pointing out is that we are now working in an over complete basis. The Pauli blocking of local spin up states on Fe sites is enforced in the global frame by constraining $\Sd_{i\uparrow}^{\dagger}$ and $\Sd_{i\downarrow}^{\dagger}$ operators to appear only in specific linear combinations that correspond to the local down spin operators. 

We obtain the spin wave dispersion by taking the second derivative with respect to $\theta$ of the total change in energy:

\begin{align} \label{eq:sw_dispersion}
  E(\mathbf q) = \frac{1}{2N}\frac{d^2}{d\theta^2} \sum_{\left| \mathbf k \right| \le k_F } \delta \epsilon(\mathbf k,\mathbf q)
\end{align}

where the sum is taken  over all occupied levels. $\delta \epsilon(\mathbf{k},\mathbf{q})$ [see Eq.~\eqref{eq:delta_e}] is the change in energy up to $\mathcal{O}(\theta^2)$ of the state in the conduction band labeled by $\mathbf{k}$. The resulting spin wave dispersion is shown in Fig.~\ref{fig:sw}. It agrees very well with the spin wave dispersion calculated by exact diagonalization on a finite lattice, also shown in Fig.~\ref{fig:sw}. No fitting parameters are used. The analytical result from the perturbative analysis, presented here, has the advantage of not being limited by finite system sizes. In paper II, we use the spin wave dispersion to extract the parameters of the effective spin Hamiltonian. 

We have further calculated the spin stiffness, $J_{\rm {eff}}=\lim_{\mathbf q \to 0} (\partial^2E(\mathbf q)/\partial q^2)$. For parameters relevant to SFMO, we find that $J_{\rm eff}=-0.035t$ which is about two orders of magnitude smaller than the band width ($W \approx 8t$) of the itinerant electrons. This justifies our initial assumption that on the time scale of the electrons, the spin waves can indeed be approximated by static spin configurations.

\section{Temperature dependent density of states} \label{sec:DOS}

We use a method that combines exact diagonalization with Monte Carlo (ED+MC) to calculate temperature dependent properties of SFMO \cite{sanyal_2009,prl_2011}. For each spin configuration, the electronic energy is calculated by exact diagonalization which is then used to update the spin configuration in the Monte Carlo algorithm. The assumption that the fast electrons relax immediately to the given spin texture has already been justified in Sec.~\ref{sec:sw} by the clear separation of time scales for the local and itinerant degrees of freedom. At each Monte Carlo step, a new random spin orientation is generated using Marsaglia's method \cite{marsaglia_1972} and acceptance is based on the Metropolis algorithm. All calculations are done on lattices up to 16$\times$16, and twisted boundary conditions are used to minimize finite-size effects. 

\begin{figure}[t]
\centering
\includegraphics[width=8.5cm]{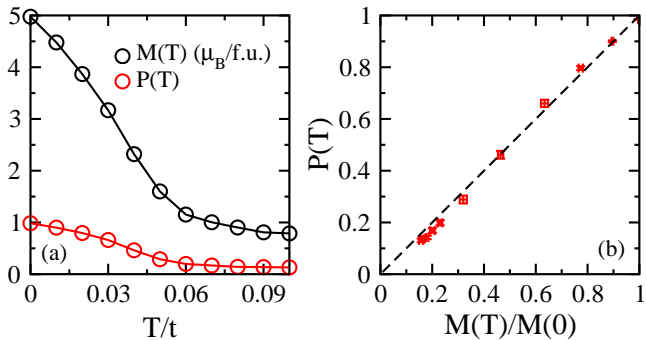}
\caption{(a) Core spin magnetization (M) and polarization of the conduction electrons at the chemical potential (P) of SFMO as a function of temperature calculated using ED+MC method on an 8$\times$8 lattice. (b) Parametric plot of P(T) against normalized M(T). It is clear that P(T) is proportional to M(T). The dashed line indicates exact proportionality.}
\label{fig:pvsm}
\end{figure}

We use ED+MC method to calculate, for the first time, the spin resolved density of states (DOS) as a function of temperature, shown in Fig.~\ref{fig:DOS}.  
At all temperatures, the spin quantization axis is defined along the direction of magnetization, which is the natural axis for a ferromagnet. 

Fig.~\ref{fig:DOS}(a) shows the spin polarized bands of the FM ground state (red indicates spin down while blue denotes spin up). The bonding band of the Fe $t_{2g\dn}$ and Mo $t_{2g\dn}$ orbitals form the conduction band while the anti-bonding band is pushed up in energy. The spin up band in the middle comes from the Mo $t_{2g\up}$ orbitals and they don't hybridize with the spin down orbitals in the perfect FM state. 

We find that for T=0, only the spin down bonding band is occupied and SFMO is a half metal, in agreement with photoemission experiment \cite{saitoh_2002} and electronic structure calculations \cite{Kobayashi,dd-sarma}.  For 0$<$T$<$\Tc, the broken time reversal symmetry leads to very different DOS for spin up and spin down sectors with the DOS at chemical potential dominated by spin down. In clear distinction from the strictly T=0 case, both spin sectors have non-zero DOS at all energies. 
Finally, for T$>$\Tc, there is no preferred spin direction and DOS for spin up and down are identical.  

As seen in Fig.~\ref{fig:DOS}(b),(c) and (d), the DOS varies smoothly with temperature. This has an important consequence that the polarization of the conduction electrons at the chemical potential, P=(N$_{\downarrow}$-N$_{\uparrow}$)/(N$_{\downarrow}$+N$_{\uparrow}$) where N$_{\sigma}$ is the density of states of spin $\sigma$ at the chemical potential, is proportional to the magnetization of Fe core spins, M, as a function of temperature \cite{prl_2011}. In Fig.~\ref{fig:pvsm}(a), M and P are plotted as a function of temperature. For better visualization of the proportionality, we have shown a parametric plot of P against a normalized M in Fig.~\ref{fig:pvsm}(b), with T as the implicit parameter. A linear fit describes the data very well. The proportionality of P and M is crucial experimentally because P(T) is the quantity of interest in spintronics applications but is difficult to measure. Our result allows direct inference of polarization from the magnetization, the latter being a much simpler quantity to measure experimentally. From a theoretical point of view also, the proportionality of P and M allows us to focus only on the magnetism. In paper II, we derive an effective classical spin Hamiltonian which describes the thermodynamics of the Fe core spins. It facilitates accurate calculation of magnetic properties and, by virtue of the proportionality, also provides realistic results for the electronic polarization.

\section{Effect of disorder}\label{sec:disorder}

\begin{figure}[t]
\centering
\includegraphics[width=8.5cm]{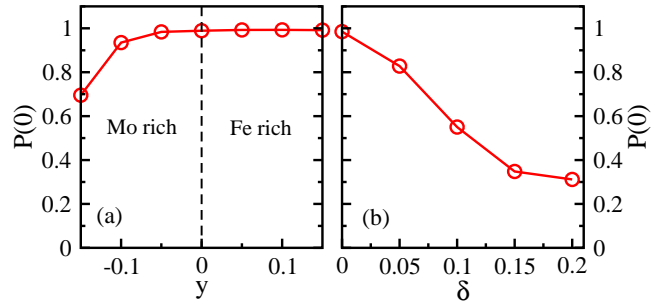}
\caption{(a) Zero temperature polarization of conduction electrons, P(0), in off-stoichiometric SFMO with general formula La$_x$Sr$_{2-x}$Fe$_{1+y}$Mo$_{1-y}$O$_6$ ($y>0$ is Fe rich while $y<0$ is Mo rich). (b) P(0) as a function of anti-site disorder. Anti-site disorder is characterized by $\delta$ which is the fraction of Fe on the wrong sublattice. Fe rich systems remain half-metallic while anti-site disorder or Mo excess rapidly decreases the polarization.}
\label{fig:disorder}
\end{figure}

In this section, we discuss the effect of disorder on electronic polarization at low T.  Its effect on magnetization is described in paper II. We consider three types of disorder: 1) excess Fe, 2) excess Mo, and 3) anti-site disorder. 

\emph{Excess Fe}: The general formula of off-stoichiometric SFMO is  La$_x$Sr$_{2-x}$Fe$_{1+y}$Mo$_{1-y}$O$_6$ and Fe rich systems correspond to $y>0$. Some Mo is substituted by Fe, which generates antiferromagnetic superexchange coupling $S(S+1)J_{SE}\approx 34 \,\mathrm{meV}$ between Fe on neighboring sites. The exchange coupling $J_{SE}$ is estimated from that of the AF insulator LaFeO$_3$ with T$_\mathrm{N}\approx 750$K using $S(S+1)J_{SE}=k_B$T$_\mathrm{N}/2$. Since Mo is also the source of itinerant electrons, excess Fe decreases the filling. In Fig.~\ref{fig:disorder}(a)  we show the electronic polarization at zero temperature, P(0), for Fe rich SFMO. An important observation is that the conduction electrons remain fully spin polarized. 
The persistent half-metallicity can be understood intuitively from the fact that at low temperatures the strong superexchange interaction locks the excess Fe on the {\em wrong sublattice} in a perfect antiferromagnetic configuration with respect to its neighboring Fe sites. Consequently, the extra Fe sites are prevented from participating in the delocalization of itinerant electrons, while the rest of the lattice continues to have a ferromagnetic ground state with spin polarized conduction electrons.

\emph{Excess Mo}: When Mo substitutes for Fe, there are regions with Fe spins that are much farther apart than in the perfect lattice. In addition, the density of carriers in Mo-rich systems is higher than that of pure SFMO. Fig.~\ref{fig:disorder}(a) shows that unlike excess Fe,
P(0) rapidly decreases as a function of excess Mo ($y<0$), consistent with DFT calculations \cite{mishra_2010}. 
The Mo rich regions constitute small regions that can be described by tight binding lattices with no preferred spin direction. They 
introduce states with both spins in the entire energy range of the conduction band
which therefore reduces the polarization at the chemical potential.

\emph{Anti-site}: Finally, anti-site disorder, which is the most common type of disorder, arises when some Fe sites exchange positions with Mo sites. It can be thought of as a combination of Fe rich and Mo rich regions while keeping the overall stoichiometry unchanged. 
Since anti-site disorder  introduces Mo rich regions, P(0) decreases with increasing anti-site disorder as shown in Fig.~\ref{fig:disorder}(b). The extent of anti-site disorder is parametrized by $\delta$ which is defined as the fraction of Fe on the wrong sublattice; complete disorder corresponds to $\delta=0.5$. While anti-site disorder behaves quantitatively like excess Mo, the electronic polarization decreases much faster with anti-site disorder. Given that fact that it is also the most common form of disorder, it is of crucial importance that anti-site disorder be minimized in order to get the high electronic polarization required for spintronics applications.

\section{Conclusion}\label{sec:conclusion}

In conclusion, we have presented here a general framework for understanding half-metallic double perovskites. While the generalized double exchange model predicts other magnetic phases as a function of Hamiltonian parameters, we have shown that for SFMO only the FM phase is relevant. 
We have calculated for the first time the temperature dependence of the spin resolved DOS and found a proportionality between the temperature dependence of the electronic polarization and the magnetization which is a significant result. It offers a much simpler method for determining the polarization. Finally, we have shown that Fe rich systems have a half-metallic ground state while anti-site disorder greatly reduces the polarization. Such understanding is crucial for spintronics applications. 

The results of paper I become the starting point for paper II. Motivated by the proportionality between the electronic polarization and the core spin magnetization, in paper II, we focus entirely on the large local spins on Fe sites and infer electronic properties from the magnetization. We derive an the effective magnetic Hamiltonian describing the thermodynamics of the classical spins. The effective Hamiltonian offers a new framework for understanding the magnetic properties in half metallic double perovskites, including the effects of disorder on the saturation value of magnetization and the \Tc. We also take advantage of the fact that Fe excess does not change the half metallic ground state to propose a novel way of increasing \Tc \, without sacrificing the polarization.

\section*{Acknowledgment}
We thank D.D. Sarma for fruitful discussions. Funding for this research was provided by the Center
for Emergent Materials at the Ohio State University, an
NSF MRSEC (Award Number DMR-0820414).


\appendix
\section{Variational Analysis}\label{appen:variational}

The electronic bands for the variational magnetic states can be obtained by Fourier transforming the double exchange Hamiltonian in Eq.~\eqref{eq:quantum_hamiltonian} into momentum space. The minimum dimension of the Hamiltonian matrix in momentum space is
related with the periodicity of the configuration. FM has periodicity of one
unit cell and every unit cell has three states, therefore $H_{FM}(\mathbf{k})$ is a 3$\times$3
matrix. However AF1 and AF2 has a minimum periodicity of two unit cell. Thus
$H_{AF1}(\mathbf{k})$ and $H_{AF2}(\mathbf{k})$ are six dimensional. The Hamiltonian matrices are
\[
H_{FM}(\mathbf{k}) =
 {\begin{pmatrix}
\Delta & 0 & -2t g_1(\mathbf{k}) \\
0 & -2t^\prime g_2(\mathbf{k}) & 0 \\
-2t g_1(\mathbf{k}) & 0 & -2t^\prime g_2(\mathbf{k})\\
\end{pmatrix} } 
\]
where 
\begin{eqnarray}
 g_1(\mathbf{k}) &=& \cos\left( \frac{k_xa + k_ya}{2} \right) + \cos \left( \frac{k_xa-k_ya}{2} \right) \nonumber \\
  g_2(\mathbf{k}) &=& \cos(k_xa) + \cos(k_ya) 
\end{eqnarray}

\[
 H_{AF1}(\mathbf{k}) =
{\begin{pmatrix}
  \Delta & 0 & \xi_k & 0 & 0 & \xi^\ast_k \\
0 & \delta_k  & 0 & \xi_k & \gamma_k & 0 \\
\xi^\ast_k & 0 & \delta_k & 0 & 0 & \gamma_k \\
0 & \xi^\ast_k & 0 & \Delta & \xi_k & 0 \\
0 & \gamma_k & 0 & \xi^\ast_k & \delta_k & 0 \\
\xi_k & 0 & \gamma_k & 0 & 0 & \delta_k
 \end{pmatrix} }
\]
\begin{eqnarray}
 \xi_k&=&-2t\cos(k_ya/2)e^{ik_xa/2}\cr
\delta_k&=&-2t^{\prime}\cos(k_ya)\cr
\gamma_k&=&-2t^{\prime}\cos(k_xa)
\end{eqnarray}

\[
 H_{AF2}(\mathbf{k})=
{\begin{pmatrix}
  \Delta & 0 & \psi_k & 0 & 0 & \nu_k \\
 0 & 0 & 0 & \nu_k & \lambda_k & 0 \\
\psi_k & 0 & 0 & 0 & 0 & \lambda_k \\
0 & \nu_k & 0 & \Delta & \psi_k & 0 \\
0 & \lambda_k & 0 & \psi_k & 0 & 0 \\
\nu_k & 0 & \lambda_k & 0 & 0 & 0 
 \end{pmatrix}}
\]
\begin{eqnarray}
 \psi_k&=&-2t\cos((k_x+k_y)a/2)\cr
\nu_k&=&-2t\cos((k_x-k_y)a/2)\cr
\lambda_k&=&-2t^{\prime}(\cos(k_xa)+\cos(k_ya))
\end{eqnarray}

The relevant energy bands are obtained by diagonalizing these matrices as a function of $\mathbf{k}$, and the energy of the conduction electrons is calculated by integrating over filled levels. The relative energy of FM, AF1 and AF2 determines the phase boundaries in Fig.~\ref{fig:phase_diagram}.  


\section{Second order perturbation theory}\label{appen:perturbation}

We describe here the perturbative calculation of the spin wave dispersion and the spin stiffness of SFMO. The Fe core spins cant by a small angle $\theta$ in a spin wave configuration [Eq.~\eqref{eq:spin_wave}], and we calculate corrections to the energy levels of the FM ground state up to $\mathcal{O}(\theta^2)$. The unperturbed Hamiltonian $H_0$ in Eq.~\eqref{eq:H_0} gives four bands with eigenvectors given by
\begin{eqnarray}
     a_1^{\dagger}(\mathbf{k}) &=& \alpha(\mathbf{k})\Sd_{\mathbf{k}\dn}^{\dagger} + \beta(\mathbf{k})c_{\mathbf{k}\dn}^{\dagger} \nonumber \\
	 a_2^{\dagger}(\mathbf{k}) &=& -\beta(\mathbf{k})\Sd_{\mathbf{k}\dn}^{\dagger} +   \alpha(\mathbf{k})c_{\mathbf{k}\dn}^{\dagger} \nonumber \\
	 a_3^{\dagger}(\mathbf{k}) &=& c_{\mathbf{k}\up}^{\dagger} \nonumber \\
	 a_4^{\dagger}(\mathbf{k}) &=& \Sd_{\mathbf{k}\up}^{\dagger} 
\end{eqnarray}    

The corresponding eigenvalues are
\begin{eqnarray}
	\epsilon_1(\mathbf k) &=& \frac{\Delta}{2} -\tp g_2(\mathbf k) - \Gamma(\mathbf{k}) \nonumber \\
	\epsilon_2(\mathbf k) &=& \frac{\Delta}{2} -\tp g_2(\mathbf k) + \Gamma(\mathbf{k}) \nonumber \\
	\epsilon_3(\mathbf{k}) &=& -2\tp g_2(\mathbf{k}) \nonumber \\
	\epsilon_4(\mathbf{k}) &=& \Delta 
\end{eqnarray}     
where
\begin{align}
  \alpha(\mathbf{k}) &= \frac{\epsilon_3(\mathbf k) - \epsilon_1(\mathbf{k})}{\sqrt{A(\mathbf{k})}} \\
  \beta(\mathbf{k}) &= \frac{2t g_1(\mathbf{k})}{\sqrt{A(\mathbf{k})}} \\
  \Gamma(\mathbf{k}) &= \sqrt{4t^2g_1^2(\mathbf k) + \left( \Delta /2  + \tp g_2(\mathbf k) \right)^2} \\
  A(\mathbf k) &= \left( \epsilon_1(\mathbf k) - \epsilon_3(\mathbf k) \right)^2 + \left( 2tg_1(\mathbf k) \right)^2 \\
  g_1(\mathbf{k}) &= \cos\left( \frac{k_xa + k_ya}{2} \right) + \cos \left( \frac{k_xa-k_ya}{2} \right) \\
  g_2(\mathbf{k}) &= \cos(k_xa) + \cos(k_ya) 
\end{align}
The first band is the bonding Fe$_\dn$-Mo$_\dn$ band. The second is the anti-bonding band. The third and fourth are the Mo$_\up$ and Fe$_\up$ bands respectively. The distance between two Fe ions is $a$. For SFMO filling only the lowest band $\epsilon_1(\mathbf{k})$ is occupied. 

\begin{figure}[t]
\centering
\includegraphics[width=8cm]{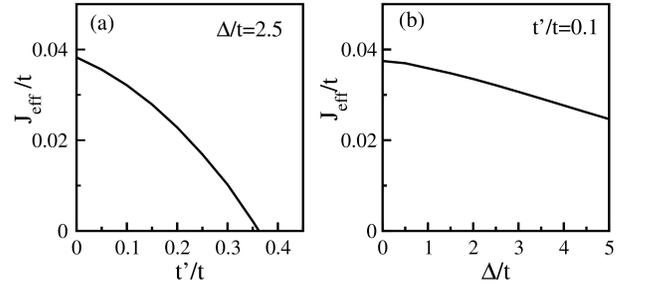}
\caption {$J_{\mathrm{eff}}$ as a function of Hamiltonian parameters: (a) $J_{\mathrm{eff}}$ vs $t^\prime$  for $\Delta =2.5t$, (b) $J_{\mathrm{eff}}$ vs $\Delta$ for $t^\prime=0.1t$. Filling is fixed at SFMO value ($n=1/3$).}
\label{fig:jeff}
\end{figure}

We next describe how the energy of the lowest band $\epsilon_1(\mathbf{k})$ is affected by $H_1$ and $H_2$. First order correction in the canting angle $\theta$ is 
\begin{equation}
\epsilon_1^{(1)}(\mathbf{k,q})=\theta \langle a_1(\mathbf{k}) \rvert H_1 \lvert a_1(\mathbf{k}) \rangle = 0
\end{equation}
The second order correction has several contributions. One of them is 
\begin{equation}
  \langle a_1(\mathbf{k}) \rvert H_2 \lvert a_1(\mathbf{k}) \rangle =  t\, g_1(\mathbf{k})\alpha(\mathbf{k})\beta(\mathbf{k})
\end{equation}
The mixing of the lowest band with the anti-bonding band gives
\begin{equation}
 \sum_{\mathbf{k^{\prime}}} \frac{\lvert \langle a_2(\mathbf{k^{\prime}}) \rvert H_1 \lvert a_1(\mathbf{k}) \rangle \rvert^2}{\epsilon_1(\mathbf{k})-\epsilon_2(\mathbf{k^{\prime}})}=0
\end{equation} 
The mixing with the Mo$\dn$ band gives 
\begin{equation}
  \sum_{\mathbf{k^{\prime}}} \frac{\lvert \langle a_3(\mathbf{k^{\prime}}) \rvert H_1 \lvert a_1(\mathbf{k}) \rangle \rvert^2}{\epsilon_1(\mathbf{k})-\epsilon_3(\mathbf{k^{\prime}})}= \frac{t^2g_1^2(\mathbf{k}-\mathbf{q})\alpha^2(\mathbf{k})}{\epsilon_1(\mathbf{k})-\epsilon_3(\mathbf{k}-\mathbf{q})}
\end{equation} 
Finally, the mixing with Fe$\up$ band gives
\begin{equation}
  \sum_{\mathbf{k^{\prime}}} \frac{\lvert \langle a_4(\mathbf{k^{\prime}}) \rvert H_1 \lvert a_1(\mathbf{k}) \rangle \rvert^2}{\epsilon_1(\mathbf{k}) - \epsilon_4(\mathbf{k^{\prime}})} = \frac{t^2g_1^2(\mathbf{k})\beta^2(\mathbf{k})}{\epsilon_1(\mathbf{k})-\Delta}
\end{equation} 
Upon collecting all the second order correction terms and simplifying them algebraically, we get the energy correction in lowest band up to $\mathcal{O}(\theta^2)$
\begin{align} \label{eq:delta_e}
  \delta \epsilon(\mathbf{k,q}) &= \epsilon_1^{(1)}(\mathbf{k,q}) + \epsilon_1^{(2)}(\mathbf{k,q}) \nonumber \\
  	&=\frac{\theta^2 t^2 g_1^2 (\mathbf k) }{ A(\mathbf k)} \left( \epsilon_3(\mathbf k) - \epsilon_1(\mathbf k ) \right)  \nonumber \\
    &+ \frac{\theta^2 t^2}{A(\mathbf k)} \frac{ g_1^2(\mathbf k - \mathbf q)\left(\epsilon_1(\mathbf k) - \epsilon_3(\mathbf k) \right)^2}{ \epsilon_1(\mathbf k) - \epsilon_3(\mathbf k - \mathbf q)}
\end{align}

Notice that the energy correction in Eq.~\eqref{eq:delta_e} has two terms. The first term comes from narrowing of conduction band in the spin wave background. It increases spin stiffness. The second term comes from hybridization of spin down conduction band with the Mo$_\up$ band and it reduces the spin stiffness. As shown in Eq.~\eqref{eq:sw_dispersion}, the spin wave dispersion can be calculated by summing $\delta \epsilon(\mathbf{k,q})$ over all filled states. 

We can also use the result from perturbative analysis to calculate the change in spin stiffness, $J_{\rm {eff}}=\lim_{\mathbf q \to 0} (\partial^2E(\mathbf q)/\partial q^2)$, as a function of Hamiltonian parameters. Fig.~\ref{fig:jeff}(a) shows the dependence of $J_{\mathrm{eff}}$ on $t^\prime$ while Fig.~\ref{fig:jeff}(b) shows how $J_{\mathrm{eff}}$ changes with $\Delta$. Increasing either $t^\prime$ or $\Delta$ decreases $J_{\mathrm{eff}}$ as expected. 



\bibliography{references}
\bibliographystyle{apsrev}

\end{document}